# Comparison of Optical, Radio, and Acoustical Detectors for Ultrahigh-Energy Neutrinos

P. B. Price

Physics Department, University of California at Berkeley, Berkeley, CA 94720, USA

**Abstract**

For electromagnetic cascades induced by electron-neutrinos in South Pole ice, the effective volume per detector element (phototube, radio antenna, or acoustic transducer) as a function of cascade energy is estimated, taking absorption and scattering into account. A comparison of the three techniques shows that the optical technique is most effective for energies below ~0.5 PeV, that the radio technique shows promise of being the most effective for higher energies, and that the acoustic method is not competitive. Due to the great transparency of ice, the event rate of AGN $\nu_e$-induced cascades is an order of magnitude greater than in water. For hard source spectra, the rate of Glashow resonance events may be much greater than the rate for non-resonant energies. The radio technique will be particularly useful in the study of Glashow events and in studies of sources with very hard energy spectra.

## 1. Introduction

Although solar neutrinos with energy up to a few MeV are being studied at several low-energy neutrino observatories, and a few neutrinos with tens of MeV were detected from supernova 1987A, the astrophysical neutrinos with energies of TeV and above that must accompany production of the highest energy cosmic rays await discovery. There is great international interest in designing and constructing an observatory optimized for mapping the sky for high-energy neutrinos from astrophysical sources such as Active Galactic Nuclei. All agree that such an observatory will require a huge volume of transparent, deep material such as ocean water or ice, which acts as both the target and the medium for detecting the charged particles produced in interactions of such neutrinos. Theoretical estimates of the low event rates of multi-TeV neutrinos originating in astrophysical sources outside the earth's atmosphere lead to the requirement of effective volumes $> 1$ km$^3$ [1].

Two first-generation instruments consisting of three-dimensional arrays of phototubes have made encouraging starts -- an array of five strings with a total of 23 working pairs of phototubes at a depth of ~1 km in Lake Baikal [2] and an array of four strings with 73 working phototubes frozen into ice at depths of 0.8 to 1 km at the South Pole [3]. It is hoped that two more arrays -- DUMAND [4] and NESTOR [5] -- can be deployed in ocean water at depths ~4 km in the next couple of years. All four instruments are designed to record arrival times and intensities of photoelectrons from Cherenkov light produced by muons or electrons created in interactions of muon- or electron-neutrinos.

A number of factors determine the cost and set a practical limit on the effective volume of a high-energy neutrino observatory. Chief among these are:

• for muon-neutrinos, ability to determine direction well enough to discriminate against downward-going neutrinos; for electron-neutrinos, ability to discriminate $\nu_e$-induced cascades against cascades induced by atmospheric muons;

• ratio of signal to noise (due to backgrounds in the medium, detector, and electronics);

• mean free paths for absorption and scattering of signals in water or ice from the products of the neutrino interaction;

• cost of equipment and cost of deploying strings at desired depths and of sending signals to surface for data analysis.

Various authors have advocated looking for acoustic signals in water or ice [6] or for radio signals in ice [7] instead of optical signals as a way of attaining a very large observatory volume at relatively low cost. Further papers on these two topics have appeared in many of the biennial proceedings of international cosmic ray conferences since the appearance of refs. 6 and 7. The purpose of this note is to compare the effective volume per detector element as a function of energy for the three types of signals. The results for observing the electromagnetic cascade resulting from interaction of an electron-neutrino in ice are presented in Fig. 1 and in Table 2 and will be discussed below. It is less straightforward to estimate $V_{eff}$ for detection of the muon into which a muon-neutrino converts, and we leave such discussion to a later paper. We briefly comment on the relative effectiveness of water and of ice for detection of electron-neutrinos. We conclude with calculations of the non-resonant and Glashow resonant event rates for the most optimistic and most pessimistic Szabo and Protheroe AGN spectra [8] and the Stecker and Salamon AGN spectrum [9].

## 2. Optical Method

To estimate the maximum effective volume $V_{eff} = 4\pi R_{max}^3/3$ for a single detector module (= phototube in its pressure vessel), consider a phototube with effective area $A_{PMT}$ at a distance r from the point where an electron-neutrino produces an electromagnetic cascade with energy $E_o$. Choose $r = R_{max}$ such that the number of photoelectrons recorded in the phototube is at least one when quantum efficiency η is taken into account and that the probability of a noise count is much less than unity for the relevant time interval for transport of the photons from the cascade to the phototube. The number of Cherenkov photons is the product of the Cherenkov yield (~170 photons/cm in the wavelength interval for greatest ice transparency, ~350 to ~500 nm) and the total shower track length for electrons + positrons in the cascade, given as (6400 m/TeV) × $E_o$ (TeV) [10]. The result is $I_o \approx 1 \times 10^8 E_o(TeV)$ photons. For neutrinos with $E_o \gg 1$ TeV, $R_{max}$ turns out to be much larger than both the absorption length, $\lambda_a$, and the scattering length, $\lambda_s$. The behavior of cascade photons is then well represented by a three-dimensional random walk with absorption [11], where $D = \frac{1}{3} c_i \lambda_e$; $\lambda_e = \lambda_s/(1 - <\cos \theta>)$; and $c_i = c/n$, the velocity of light in ice. The light intensity at a distance r away from the source and at time t is given by

$$f(r,t) = \frac{1}{(4\pi Dt)^{3/2}} \exp(-r^2/4Dt) \exp(-c_i t/\lambda_a) \quad (1)$$

Integrating over all times gives for the density of photons reaching a distance r

$$I = I_o \exp(-\alpha r)/4\pi Dr \quad \text{photons } (m^3/s)^{-1} \quad (2)$$

The flux at distance r is given by

$$F = I c_i /4 = I_o c_i \exp(-\alpha r)/16\pi Dr \quad \text{photons } m^{-2} \quad (3)$$

where $\alpha = (3/\lambda_e \lambda_a)^{1/2}$.

For a 3-inch phototube with $\eta \approx 25\%$ and $A_{PMT} = 44$ cm$^2$, the noise rate at -50° C is only ~100 Hz, which produces a negligible background within the time window of ~10 µs for collection of all diffusing photons from a cascade. (The advantage of this small phototube over the 8- to 16-inch phototubes used in experiments at lower energy is the low cost of the tube, pressure vessel, and drilling operation.) The criterion for finding $R_{max}$ is then:

$$F A_{PMT} \eta = I_o c_i \exp(-\alpha R_{max})/16\pi D R_{max} A_{PMT} \eta \geq 1 \quad (4)$$

Two of the solid curves for optical detection in Fig. 1 are estimates of $V_{eff}$ for "ice, bubbles", at a depth of 900 m at South Pole (with $\lambda_a = 150$ m, $\lambda_s = 0.13$ m [11]) and "ice, no bubbles", at a depth of 1600 m, where bubbles are expected to be absent [12]. In the latter case, scattering is from dust in the ice [13]. The mean value of $\cos \theta$ increases from ~0.75 for smooth, spherical bubbles [11] to ~0.9 for dust in ice [14]. The value of $\lambda_a$ is the harmonic sum of the intrinsic absorption by perfect ice and the absorption by dust, taken to be $\lambda_a(dust) = 8 \times \lambda_s(dust)$ [14].

Since F is proportional to $I_o (A_{PMT} \eta)^{-1}$ and thus to $E_o (A_{PMT} \eta)^{-1}$, it is easy to generate curves for different size phototubes and for different numbers of photoelectrons by simply translating a curve along the energy axis in Fig. 1. (At a fixed value of $V_{eff}$ or

$R_{max}$, the cascade energy $E_o$ corresponding to a given mean number of photons reaching the PMT scales as $(A_{PMT} \eta)^{-1}$.)

## 3. Radio Method

For radio wavelengths greater than the cascade dimensions, the radio power in Cherenkov radiation is coherent and scales with the square of the primary cascade energy and inversely with the square of the radiation length of the medium. The mean Cherenkov angle is ~56°, with a frequency-dependent Gaussian width of ~2.4°(500 MHz/ν) [10]. Ice has enormous advantages over air and water: the radiation length in ice is ~$10^{-3}$ that in air, and the absorption length of cold ice is ~$10^3$ that of water for 500 MHz radio signals. For ice at 500 MHz $\lambda_a$ grows from ~0.1 km at -5° C to ~ 1.2 km at -50° C. The sensitivity peaks in the frequency region ~100 to ~400 MHz.

Frichter et al. [15] have recently estimated the sensitivity of a biconical antenna in deep Antarctic ice to radio signals generated in interactions of neutrinos from active galactic nuclei. They capitalized on earlier results of Zas et al. [10] for radio field intensity as a function of frequency and angle with respect to cascade direction. They showed that the event rate integrated over all angles goes through a gentle maximum for an antenna situated at a depth of ~600 m. Two factors lead to the maximum. Because of absorption in the earth, the flux in deep ice of downward-going neutrinos of very high energies is far higher than of upward-going neutrinos, as a consequence of which the number of detectable cascades is increased by increasing the thickness of ice above an antenna. Opposing this is the increase in ice temperature with depth, which decreases the absorption length and reduces $R_{max}$. They calculated $V_{eff}$, the effective volume, as a function of cascade energy and nadir angle for a single antenna with a signal to noise ratio S/N ≥ 1 at a depth of 600 m, taking into account the probability of absorption of the parent neutrino in passing through part of the earth.

With one modification of the results of Frichter et al., the dashed curve in Fig. 1 gives $V_{eff}$ at a depth of 600 m as a function of cascade energy for a biconical radio antenna of half-cone height 11 cm and half-angle 30°, the size chosen by them as optimal for the frequency band 100 to 1000 MHz. They overestimated the absorption length, because they assumed the same temperature vs depth measured at Vostok Station, which is about 8° colder than expected for South Pole [13]. The present paper employs a more accurate absorption length vs depth based on measurements of temperature at depths down to 1 km at South Pole and a thermal model for greater depths. Note that at low neutrino energies and correspondingly small effective volumes that satisfy the criterion S/N ≥ 1, absorption of radio signals is negligible, and $V_{eff}$ is proportional to $E^3$, since the volume enclosed by two concentric cones with mean angle of 56° and finite angular thickness ~2.4°(500 MHz/$f$) grows as the cube of the radius.

In considering construction of a large array, one needs to consider distortion of trajectories of radio signals as a function of depth. Near the surface, due to the dependence of refractive index on ice density, radio signals follow curved trajectories and may even undergo total internal reflection; at great depths, signals can reflect from the bedrock. The ice density monotonically increases with depth to a constant value of 0.92 g cm$^{-3}$ at depths greater than ~150 m. For a string of antennas at greater depth, the Cherenkov cone of a downward going cascade starting in the top 150 m will be distorted, whereas for an upward going cascade this should not be a problem. For a string of radio antennas extending to a depth of 1300 m, too little power would be reflected from bedrock at a depth of ~2900 m to cause a problem. The attenuation is given by
I(1300)/I(2900) = exp(-∫sec θ dz / $\lambda_a$(T)), where $\lambda_a$(T) is obtained by using data for absorption as a function of temperature [16] and a model for the change of ice temperature with depth. The attenuation for an antenna at 1300 m is ~$10^{-3}$ in the worst case of vertical reflection. Curvature due to the dependence of refractive index on temperature is negligible: $n^{-1} \partial n/\partial T \approx 2 \times 10^{-4}$ per Kelvin or ~0.5% for a change of 30 K.

## 4. Acoustical Method

In contrast to optical and radio Cherenkov signals, the mechanism of production of an acoustic signal is believed to be thermoelastic: the medium suddenly expands when heated by the energy deposited by the cascade. The amplitude of the bimodal pressure pulse (compression followed by rarefaction) at a given distance increases linearly with energy deposited, decreases as (diameter of cascade)$^{-2}$, and increases as the Grüneisen parameter $\beta v_L^2/C_p$, where $\beta$ = volume expansivity, $v_L$ = velocity of longitudinal wave, and $C_p$ = specific heat at constant pressure. The Grüneisen parameter is an order of magnitude greater for ice at -50° C than for water. The acoustic pulse is modeled by regarding the cascade as a cylindrical antenna of radius $b \approx 2$ cm and length $y \approx 5$ m, the values depending on the particular cascade model. For water or ice with critical energy $E_{crit} = 73$ MeV and radiation length $X_o = 36$ g cm$^{-2}$ the length $y \approx 3X_o(\ln(E_o/E_{crit}))^{1/2} \approx 5$ m, almost independent of cascade energy over the interval of interest, 0.1 to 10 PeV. Likewise, one can assume b to be constant over the same interval. In the near field the acoustic wave expands normal to the cascade axis as a disc of nearly uniform thickness $h \approx 2$ y. The transition from near field to far field occurs at ~200 m, and the peak amplitude as a function of lateral distance falls off only as $r^{-1/2}$ for shorter distances. As cascade theory has improved, the estimated peak pressure at a given cascade energy has increased. For a 10 PeV cascade in sea water at a distance of 400 m, Learned [17] predicted a peak pressure of ~4.5 µPa; Askaryan et al. [7] predicted ~27 µPa; and Dedenko et al. [18] recently predicted ~60 µPa. The larger peak pressure estimated in [18] is mainly due to the narrower radial distribution b of electrons in the cascade model in comparison with models of previous authors. In addition to the timing information received at different locations, a measurement of the ratio of the compression peak to the rarefaction peak would give an estimate of distance [7].

For consistency with the criterion (S/N ≥ 1) for estimating $V_{eff}$ for radio signals, $V_{eff}$ is defined as the volume of ice within which S/N ≥ 1. For ice at -50° C, $v_L$ = 3900 m s$^{-1}$, and noise equivalent pressure = $(4\pi kT f^2 \delta f/v_L^2)^{1/2} \approx 10^{-4} f \delta f^{1/2} \approx 270$ µPa with $f_{peak}$ = 20 kHz and $\delta f \approx f_{peak}$. $V_{eff} = \pi R_{max}^2 h$, where h ≈ 10 m for $E_o$ = 10 PeV and $R_{max}$ is the disc radius at which S/N drops to 1. I used the calculated peak pressure as a function of radial distance in [7] but scaled up by a factor 10 to convert from sea water to ice and by another factor of 2.2 to reflect the cascade model of [18].

A thorough discussion of absorption and scattering of acoustic waves in ice is given in [19]. Scattering and absorption by bubbles and dust are negligible. Rayleigh scattering from ice crystal boundaries leads to a mean free path

$$\lambda_s = 8.6 \text{ km } (0.2 \text{ cm}/a)^3 (20 \text{ kHz}/f)^4 \quad (5)$$

where $a$ = mean crystal size. For $a$ estimated as ~0.2 cm and $f$ < 25 kHz, the value of $\lambda_s$ is greater than 3.5 km.

The main contribution to absorption is relaxation due to molecular reorientation, with an absorption mean free path $\lambda_a = v_L/f\delta$, where the logarithmic decrement is given by

$$\delta = 4\pi \delta_m f \tau_m / (1 + 4\pi^2 f^2 \tau_m^2) \quad (6)$$

with $\delta_m$ experimentally determined to be ~0.025 for propagation of a longitudinal wave in the basal plane of ice, and $\tau_m = \tau_o \exp(U/kT)$, with activation energy U = 0.58 eV. At a temperature of -45° C, corresponding to a depth of 1.3 km, the value of $\lambda_a$ obtained from [19] is 1 km. At shallower depths where the temperature is ~-50° C the value of $\lambda_a$ is 2 km.

The dotted curve in Fig. 1 gives $V_{eff}$ for acoustic signals, assuming $\lambda_a$ = 2 km and $\lambda_s$ = 5 km (applicable to a range of depths down to ~1 km). Absorption causes the curve to bend over for large $E_o$ (and thus large $V_{eff}$).

## 5. Relative Merits of the Three Techniques

With the information in Fig. 1 and Table 1, we can compare the relative merits of the three techniques as components of a large array. Although the costs per module appear grossly comparable, the optical technique has great advantages over the others. Electronics and computational support for signal readout and analysis have already been developed and muon signals are routinely detected in both water and ice, whereas such is not the case for the other two techniques. Small phototubes have proven reliability, low cost, high quantum efficiency, sensitivity to individual photoelectrons, and low probability of a background count during the time interval of ~$10^{-5}$ sec for diffusion of photons from a cascade. At energies well below 1 PeV a simple 3-inch phototube in an inexpensive pressure module clearly wins out over the other two techniques.

Above an energy of about 1 PeV the effective volume for a radio antenna exceeds that for a 3-inch phototube, the advantage growing with increasing energy. Since absorption and scattering are far smaller for a radio signal than for an optical signal, it should be possible to reconstruct the Cherenkov cone of the electromagnetic cascade with relatively little distortion and thus to reconstruct the trajectory of the parent $\nu_e$ from the arrival times of radio signals at the antennas of an array. In contrast, it will be difficult to infer direction of a $\nu_e$ using optical signals in an array without reducing the potentially large effective volume per element, since at distances from the vertex large with respect to $\lambda_s$ the photons will diffuse with spherical symmetry. The main concerns with the radio technique are that radio noise in deep ice has not yet been studied, receiver response to known calibration signals has not yet been determined, and the event rate for PeV neutrinos may be extremely low. To provide a convincing test of the radio technique it will be necessary first to perform *in-situ* tests of ambient noise and of antenna sensitivity to known calibration signals and then to build and test a sizable array.

From the curves in Fig. 1 it should be clear that the acoustic technique is far too insensitive to compete with the other two. The efficiency for conversion of thermal energy into sound is extremely low. Even if ambient noise in the ice were to prove to be acceptably low, thermal noise in a pressure transducer places stringent requirements on the minimum cascade energy for which S/N ≥ 1. Reducing noise by reducing bandwidth to less than about 20 kHz would not help because it would cut out useful signal. Further, because the thickness of the disc-shaped pressure wave is only ~10 m thick, a large effective volume implies a large disc radius, so that absorption in the plane of the disc restricts the values of $V_{eff}$ for multi-PeV cascades.

## 6. Practical Considerations

The Cherenkov cones for optical and radio emission are peaked at ~41º and ~56º respectively, and the disc-shaped acoustic signal is accurately normal to the cascade axis. Thus, in principle, not only the vertex and energy of an electromagnetic cascade but also its direction can be determined, if the spacing of detector elements of an array is not greater than $\lambda_s$. The large values of $\lambda_s$ for the radio and acoustical techniques lead to little distortion of directional information even for spacings between elements of as large as 1 km. In contrast, with an expected value $\lambda_s \approx 20$ m for optical photons in dusty ice, there are two options. If the elements are placed far apart in order to maximize $V_{eff}$, the photons diffuse outward from the cascade as if from a point source, without conveying directional information. If directional information is required, the phototubes must be spaced a distance not much greater than $\lambda_s$ and the photon flux at the phototube must be great enough to be able to select the few photons that arrive without scatters (the "leading edge" time [20]).

To extract all the information about a cascade including its direction can best be done by measuring signal arrival times at seven or more detectors in order to determine the seven cascade parameters ϕ, θ, x, y, z, time, and energy. To do this the array must be spaced densely enough that at least seven detectors will sense the signal. Except for a small fraction of phase space corresponding to cascades in special orientations of high symmetry, this criterion can be taken into account by having at least seven elements within each volume $V_{eff}$. To allow for fluctuations in number of optical photons or in S/N, a more conservative criterion such as $V_{eff}/20$ might be advisable. Recognizing that the hole-drilling operation adds considerably to the cost of an array, a horizontal spacing between strings of detectors that is considerably larger than the vertical spacing between detectors on a string is favored.

To avoid bubbles, which reduce $\lambda_s$ in the top 1.5 km, and to avoid the region of large shear rate in the warm ice near bedrock, strings in an optical array could extend from depths of 1.5 to 2.3 km. To avoid trajectory curvature due to the density gradient in the upper 0.15 km and to avoid the region of reduced $\lambda_a$ in the deep warm ice, strings of radio antennas could occupy depths of 0.3 to 1.3 km. The optimal array configuration for detection of cascades induced by astrophysical electron-neutrinos up to energies of many PeV might utilize both optical and radio elements. It is premature to speculate on the design of a giant array until tests of the radio technique have been carried out.

For both the optical and radio techniques one must contend with a potentially serious background due to bremsstrahlung and pair production by downward going high-energy muons that originate in high-energy cosmic ray interactions in the atmosphere above the South Pole. The main uncertainty is the magnitude of the contribution to muons due to prompt decay of charmed particles [19, 20]. In a comparison of five models of charm cross sections, Zas et al. [21] found very large variations in the predicted spectra of prompt muons. More recently, with an improved treatment of charm particle production, Gondolo et al. [22] predicted much lower prompt fluxes, with correspondingly better prospects for detecting muons from astrophysical neutrino sources. They found that the predicted diffuse neutrino fluxes from active galactic nuclei [21,23] exceed their predicted atmospheric neutrino background at all energies above ~0.1 PeV.

With the optical technique it may in principle be possible to reject the background of cascades induced by atmospheric muons by using the arrival times of "leading edge" photons from relatively closely spaced mini-cascades along the high-energy muon track to distinguish the muon as a line source from a single $\nu_e$-induced cascade concentrated in a region a few meters long [20]. Because of the lower sensitivity of the radio technique it will be more difficult to distinguish a $\nu_e$-induced cascade from the signal of a high-energy muon that happens to lose a large fraction of its total energy in a single cascade, since radio signals from other points along the muon's trajectory may be undetectably weak.

## 7. Comment about Underwater Arrays

The strong absorption of radio signals in water ($\lambda_a \approx 50$ cm at a frequency of 500 MHz) eliminates the radio technique from consideration. The Grüneisen parameter for water is only about one-tenth as large as for ice, which makes the acoustic technique relatively less attractive for water than for ice. The absorption length $\lambda_a$ for optical signals is ~40 m at a wavelength of ~400 nm for the DUMAND site (Pacific Ocean floor near Hawaii) and 55 ± 10 m for the Nestor site (Mediterranean Sea floor near Pylos) [24]. Measurements for Lake Baikal [25] give $\lambda_a = 20.5 \pm 2$ m and $\lambda_s \approx 10$ m at a wavelength of ~480 nm. The solid curve in Fig. 1 labeled "water", based on the values $\lambda_a = 20.5 \pm 2$ m, $\lambda_s \approx 10$ m, shows that the shorter absorption and scattering lengths lead one to expect arrays to have much smaller volume in water than in bubble-free South Pole ice, but larger volume than in bubbly ice at depths less than ~1 km.

## 8. Event Rates for AGN Electron-Neutrino Energy Spectra

To compare the effectiveness of the optical and radio techniques, I have calculated the event rates per year for the AGN energy spectra of Stecker and Salamon [9] and for the most optimistic and least optimistic of the AGN spectra of Szabo and Protheroe [8], assuming $\nu_e + \bar{\nu}_e$ fluxes equal to one-half of $\nu_\mu + \bar{\nu}_\mu$ fluxes. The event rates per year per

TeV with values of $V_{eff}$ taken from Fig. 1 are shown in Fig. 2. The integrated event rates per year for resonant $\bar{\nu}_e$ (Glashow) and non-resonant $\nu_e + \bar{\nu}_e$ are given in Table 2. The resonant cross section per electron at 6.4 PeV is $\sim 4.7 \times 10^{-31}$ cm$^2$ for the sum of the three leptonic and six hadronic W$^-$ decay channels, and the integrated rate per $\bar{\nu}_e$ is $2.4 \times 10^{-25}$ cm$^2$. Note that although the caption to Table 2 indicates rate per detector element, it should be stressed that more than one element is required for analysis of a cascade.

## 9. Conclusions and Comments

Due to the great transparency of ice, the event rate of AGN $\nu_e$-induced cascades is an order of magnitude greater than in water. The event rate for Glashow resonance events will dominate over the rate for non-resonant energies if the $\nu_e + \bar{\nu}_e$ source spectrum is hard out to multi-PeV energies. Thus, the rate for the resonant process dominates for the spectrum of Stecker and Salamon [9] and for radio detection of the most optimistic spectrum of Szabo and Protheroe [8], but is completely negligible for their most pessimistic spectrum and for the spectrum of Sikora and Begelman [23], which cuts off at an energy well below the resonant energy of 6.3 PeV. For the most pessimistic spectrum of Szabo and Protheroe, the optical technique is an order of magnitude more sensitive than the radio technique. A combination of the optical and radio techniques might make it possible to distinguish among predicted AGN spectra of various hardnesses.

The optical technique is highly developed, most sensitive, and unchallenged at energies well below 1 PeV. The radio technique is theoretically well founded but needs to be calibrated with a radio transmitter of known intensity in deep ice and measurements made of ambient radio noise *in situ*. It is potentially attractive at energies above ~1 PeV and for relatively undistorted imaging of cascades with a large array. The acoustical technique is least sensitive, the mechanism of energy conversion from a cascade to an acoustic signal is very inefficient, and no tests have been carried out with particle beams in ice.

**Acknowledgments**

I am indebted to Steve Barwick, George Frichter and Yudong He for useful discussions. This research was supported in part by National Science Foundation grant PHY-9307420.

**References**

[1] S. Barwick, F. Halzen, D. Lowder, T. Miller, R. Morse, P. B. Price, and A. Westphal, J. Phys. G: Nucl. Part. Phys. 18 (1992) 225.

[2] I. A. Belolaptikov et al., Proc. 24th Inter. Cosmic Ray Conf. Vol. 1, p. 742 (Rome, 1995).

[3] See papers by AMANDA Collaboration in Proc. 24th Inter. Cosmic Ray Conf. (Rome, 1995).

[4] P. K. F. Grieder, talk given at "Trends in Astroparticle Physics," in Nucl. Phys. B (Proc. Suppl.) 43 (1995) 145.

[5] NESTOR Collaboration, in Proc. 24th Cosmic Ray Conf., Vol. 1, p. 1080 (Rome, 1995).

[6] G. A. Askaryan, Sov. Phys. JETP 14 (1962) 441.

[7] G. A. Askaryan et al., Nucl. Instr. Meth. 164 (1979) 267.

[8] A. P. Szabo and R. J. Protheroe, Astropart. Phys. 2 (1994) 375.

[9] F. W. Stecker and M. H. Salamon, Space Science Reviews, in press (1995).

[10] E. Zas, F. Halzen, and T. Stanev, Phys. Rev. D 45 (1992) 362.

[11] P. Askebjer et al., Science 267 (1995) 1147.

[12] P. B. Price, Science 267 (1995) 1802.

[13] P. B. Price et al., J. Glaciology, in press (1995).

[14] L. Bergström et al., to be published (1995).

[15] G. M. Frichter, J. P. Ralston, and D. W. McKay, submitted to Phys. Rev. D (1995).

[16] J. Westphal, unpublished results, cited in S. G. Warren, Appl. Opt. 23 (1984) 1206.

[17] J. G. Learned, Phys. Rev. D 19 (1979) 3293.

[18] L. G. Dedenko, S. Kh. Karaevsky, A. A. Mironovich, and I. M. Zheleznykh, Proc. 24th Inter. Cosmic Ray Conf., Vol. 1, p. 797 (Rome, 1995).

[19] P. B. Price, Nucl. Instr. Meth. A325 (1993) 346.

[20] P. C. Mock et al., Proc. 24th Inter. Cosmic Ray Conf., Vol. 1, p. 758 (Rome, 1995).

[21] E. Zas, F. Halzen, and R. A. Vazquez, Astroparticle Phys. 1 (1993) 297.

[22] P. Gondolo, G. Ingelman, and M. Thunman, to be published (1995).

[23] M. Sikora and M. Begelman, in *Proc. Workshop on High Energy Neutrino Astrophysics*, Honolulu, eds. V. J. Stenger et al. (World Scientific, Singapore, 1992), p. 114.

[24] E. G. Anassontzis, P. Ioannou, Chr. Kourkoumelis, L. K. Resvanis, and H. Bradner, Nucl. Instr. Meth. A 349 (1994) 242.

[25] I. A. Belolaptikov et al., Proc. 24th Inter. Cosmic Ray Conf. Vol. 1, p. 1043 (Rome, 1995).

Table 1. Signals of PeV $\nu_e$-Induced Cascades in Ice at -50°C

|  | Optical (dust, not bubbles) | Radio | Acoustic |
|---|---|---|---|
| wavelength (cm) | (3 to 6) x $10^{-4}$ | 30 to 300 | 10 to 80 |
| absorption length (km) | 0.12 | 1.2 | 2 |
| scattering length (km) | 0.02 | >2 | 5 |
| optimum depth (km) | 1.5 to 2.3 | 0.3 to 1.3 | 0.3 to 1.3 |
| cost per element (not including cable) | ~$500 (3-inch PMT) | ~$2000 (dipole) | ~$1000 |
| emission relative to cascade | 41° cone but diffusing | 56° cone (2-3° half-angle) | thin disc $\perp$ to cascade axis |

Table 2. Event Rates per Year for a Single Element

|  | Szabo and Protheroe (max.) | | | Sz. and Proth. (min.) | | | Stecker and Salamon | | |
|---|---|---|---|---|---|---|---|---|---|
|  | non-res. | reson. | total | non-res. | reson. | total | non-res. | reson. | total |
| optical, ice | 160 | 185 | 345 | 35 | .01 | 35 | 20 | 185 | 205 |
| optical, water | 10 | 10 | 20 | 4 | .0005 | 4 | 1.5 | 10 | 11.5 |
| radio, ice | 110 | 950 | 1060 | 4 | 0.05 | 4 | 50 | 950 | 1000 |

**Figure Captions**

1. Effective volume per detector element as a function of the energy of a $\nu_e$-induced cascade for visible light in a 3-inch phototube, for radio signals in a bi-conical antenna limited by thermal noise to S/N = 1, and for acoustic signals in a transducer limited by thermal noise to S/N = 1. Note that "ice, no bubbles" applies to depths 1.5 to 2.3 km; "ice, bubbles" applies to a depth of 0.9 km; "water" is calculated for values of $\lambda_a$ and $\lambda_s$ measured in Lake Baikal at depth 1 km; and all three are for a hypothetical 3-inch phototube.

2. Predicted event rates for the values of $V_{eff}$ given in Figure 1, with Glashow resonance contribution to $\bar{\nu}_e$ included: (a) Most optimistic $\nu_e + \bar{\nu}_e$ spectrum of Szabo and Protheroe [8], (b) most pessimistic $\nu_e + \bar{\nu}_e$ spectrum of Szabo and Protheroe, (c) $\nu_e + \bar{\nu}_e$ spectrum of Stecker and Salamon [9].

Figure 1

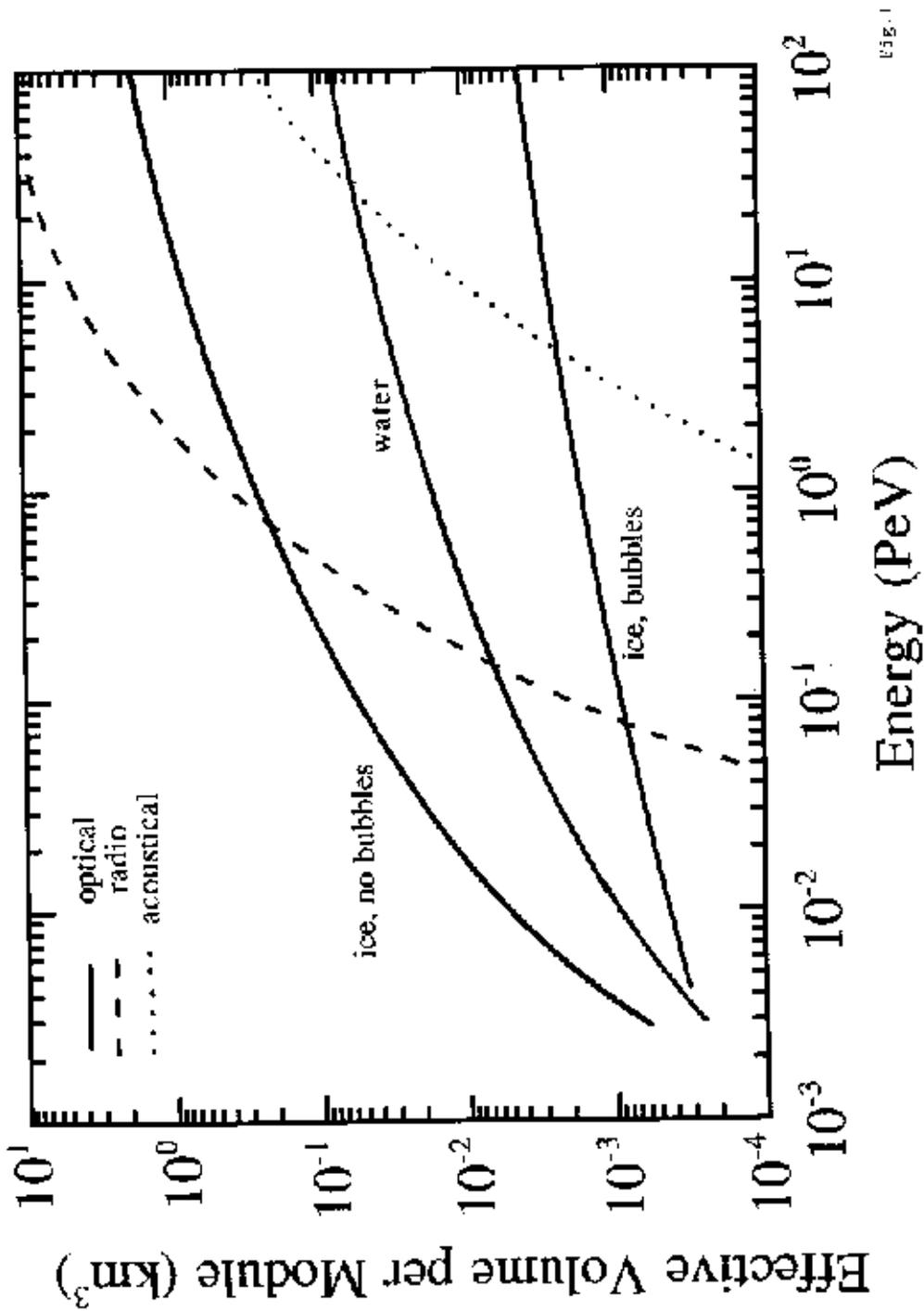

Fig. 1

Figure 2(a)

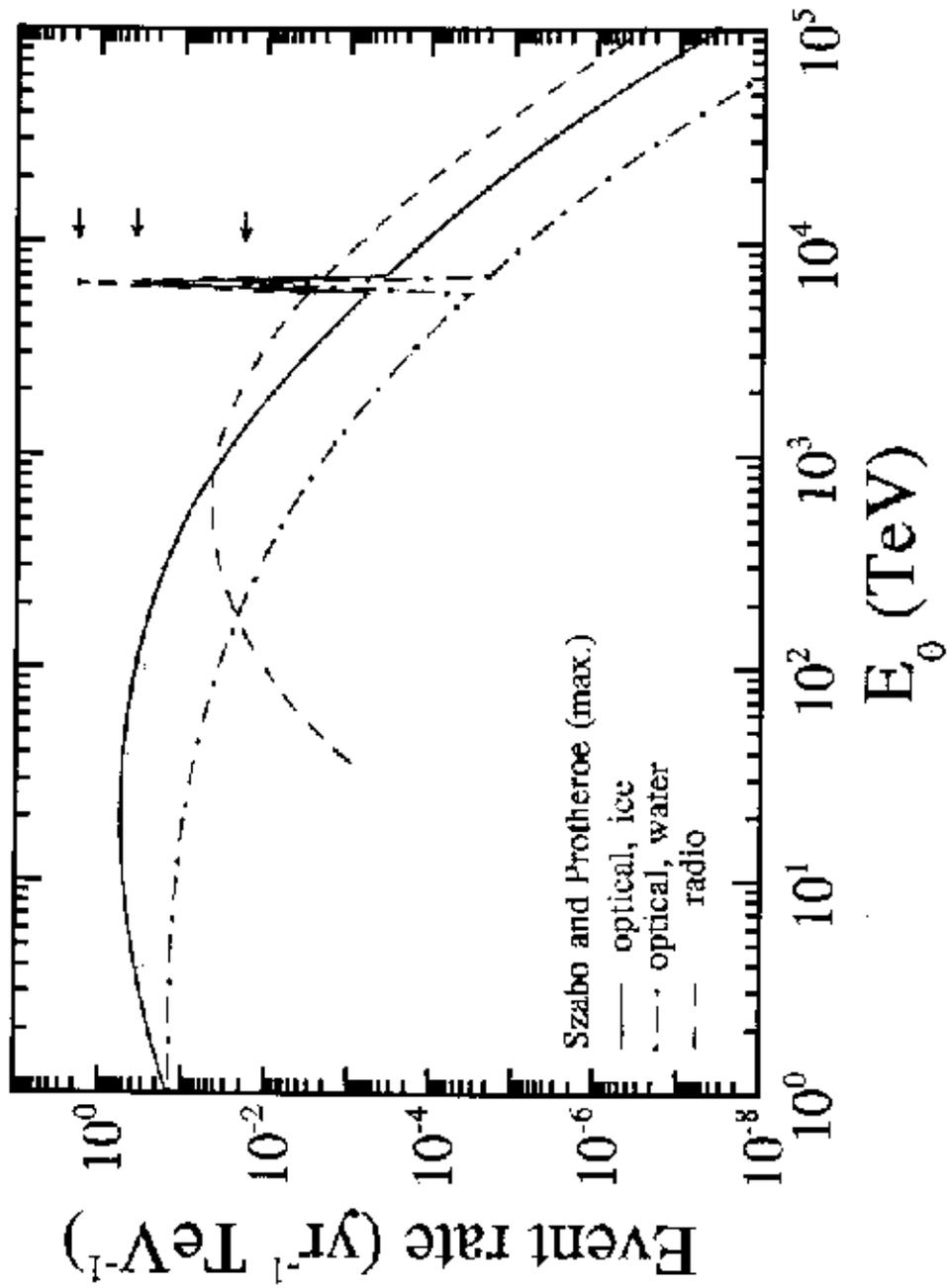

Figure 2(b)

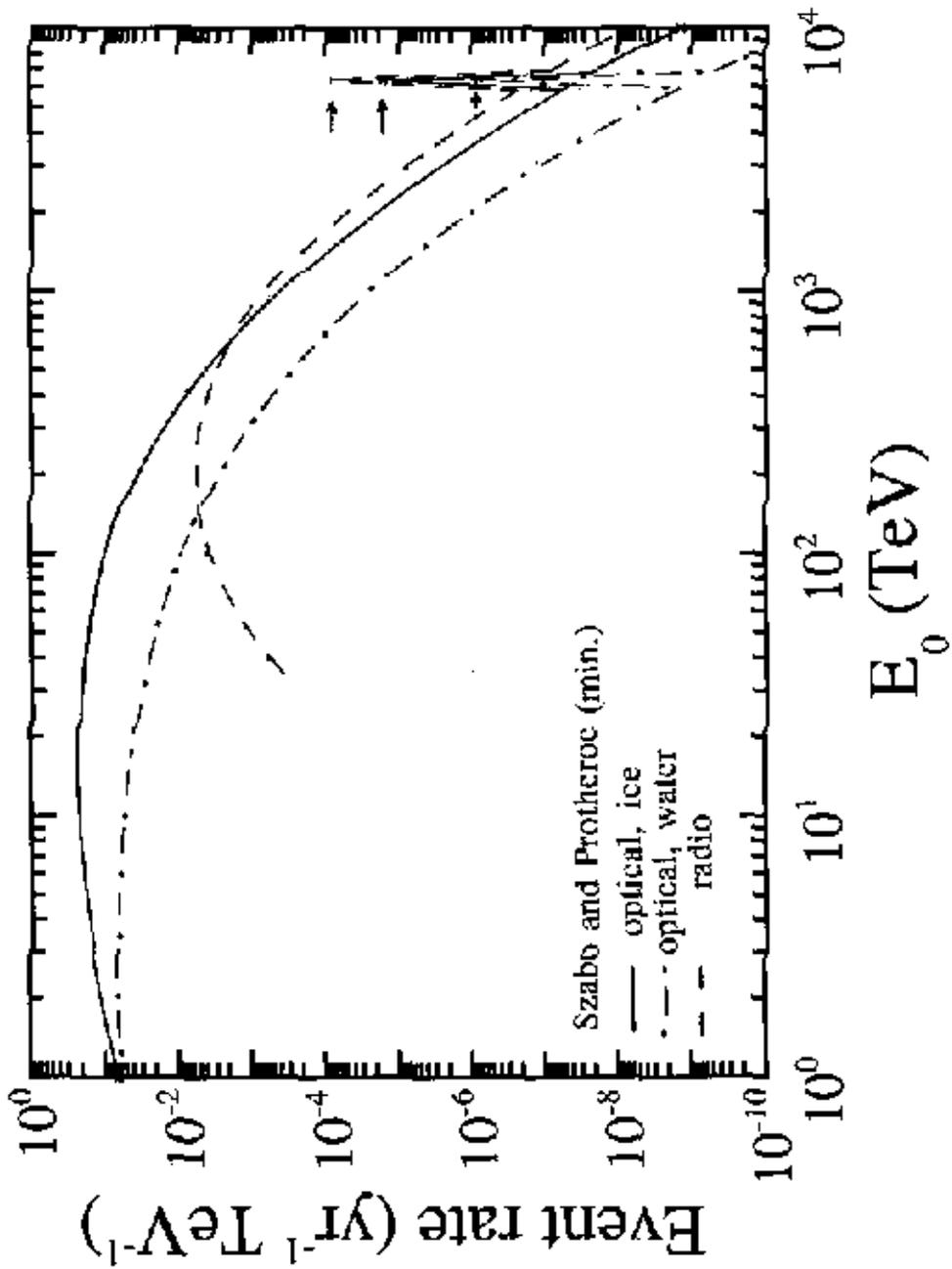

Figure 2(c)

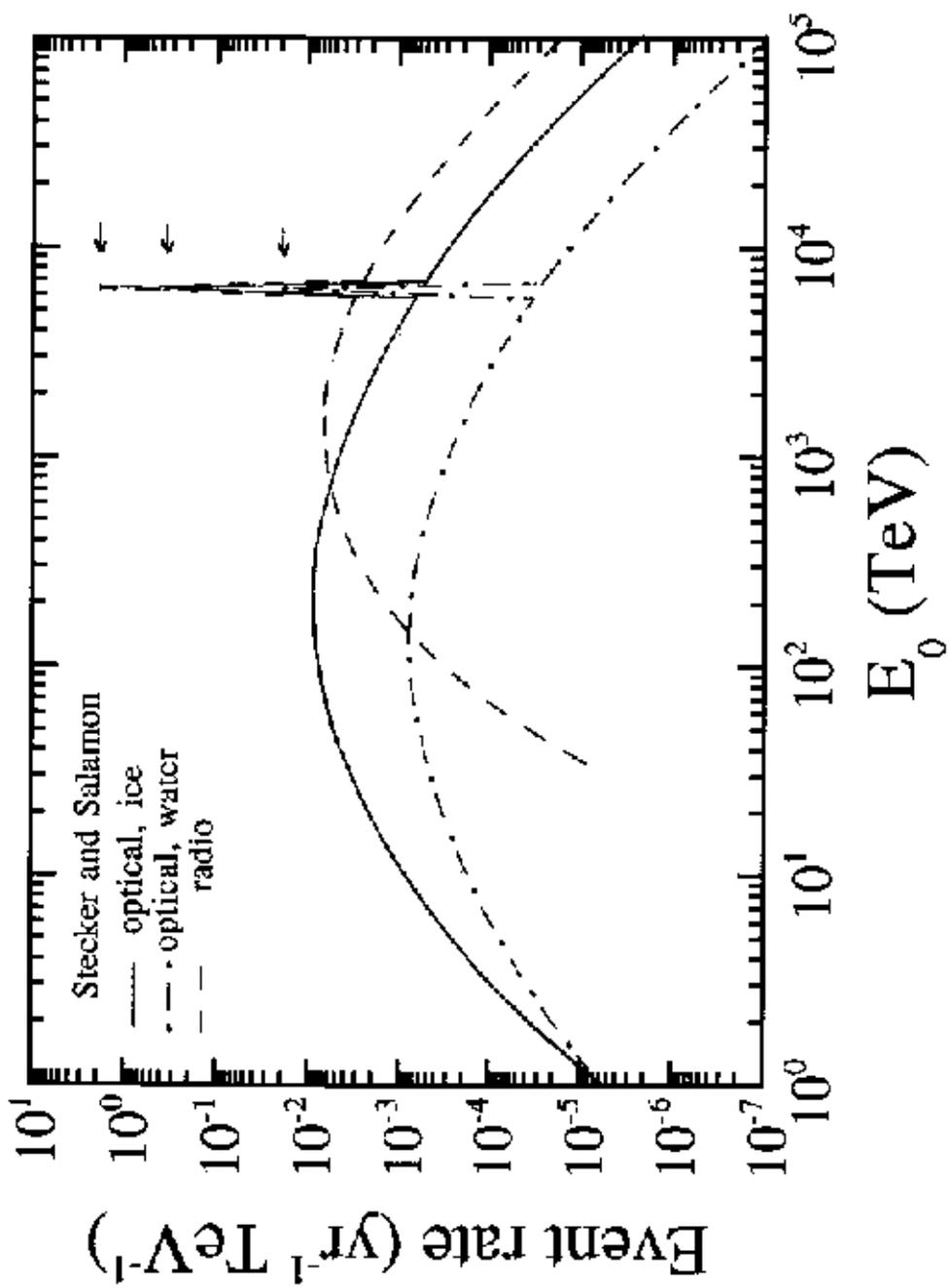